\documentclass[12pt]{iopart}
\usepackage{iopams,graphicx,bm,color}
\newcommand{\bracket}[1]{\left\langle #1\right\rangle}

\newcommand{\cotanh}{\mathrm{cotanh}}
\newcommand{\hu}{\hat{	u}}

\newcommand{\hz}{\hat{z}}

\newcommand{\tc}{\hat{t}}

\begin{document}
\title{The Storm and Nelson's model for polymer stretching revisited}
\author{Francesco A. Massucci$^1$, Isaac P\'erez Castillo$^1$, Conrad P\'erez Vicente$^2$}
\address{$^1$ Department of Mathematics, King's College London, Strand, London WC2R 2LS, United Kingdom}
\address{$^2$ Department de Fisica Fonamental, Universitat de Barcelona, Diagonal 647, 08028 Barcelona, Spain}
\date{\today}

\begin{abstract}
The technological innovation of the experimental techniques involved in molecular biology allows nowadays to perform experiments almost unimaginable a few decades ago. Among these a particularly interesting role is played by single molecule manipulation experiments, aiming at understanding the mechanical properties of the fundamental bricks of life. DNA stretching experiments show for instance that both single stranded and double stranded DNA (respectively ss-DNA and ds-DNA) have first an elastic elongation, due to entropic effects, follow by an enthalpic regime in which the molecule's constituents themselves start to elongate. Moreover, when the stretching force reaches values around 65 pN ds-DNA can suffer a second type of transition, after which the molecule reaches an elongation of about 1.7 times its contour length.\\ 
Besides those fascinating experiments it is necessary to formulate mathematical models to give a quantitative flavour to the already large amount of experimental results. These models need to be realistic enough to describe the complexity observed in the experiments yet sufficiently simple to be conducive to fitting the experimental data. So far all the models proposed to reproduce experimental measurements of DNA elongation failed to satisfy both grounds, either lacking some fundamental properties of the molecule or being too involved to analyse experimental data.\\
In this paper we reconsider a model proposed a few years ago by Storm and Nelson \cite{storm-nelson} which represented an interesting step towards a theory that could synthesise the properties of accuracy and simplicity. We  solve the model using the cavity method, avoiding any mathematical approximation, yielding results that not only fit successfully experimental measurements performed on single and double stranded DNA but also produce sensible physical values of the parameters of the model. Furthermore, to confirm the mathematical quality of the approach, our findings are also compared to Monte Carlo simulations.
\end{abstract}

\maketitle 

\section{Introduction}
The impressive development of experimental techniques in molecular biology made possible a brand new class of experiments, among which a particularly important role is played by the ones of single molecule manipulation. Such experiments aim to understand the response of the very fundamental bricks of living beings to a mechanical stimulus (e.g. DNA pulling experiments can give some insights on the mechanical behaviour of the Nucleic Acid during replication).\\
The pioneering works of Bustamante \cite{bustamante1992} in the early 90s opened a research field that uncovered many peculiar features of biological molecules under tension, showing, for example, that DNA and RNA have an elastic response to small pulling forces and  a resistance to bending \cite{elasticity,bendingstiff}, implying that the correlation along the molecular chain is non negligible.\\
Some of these experiments showed as well another interesting property of double stranded DNA (dsDNA): it has been seen that when B-DNA (i.e. the DNA in its well-known helical configuration) is stretched by  forces of around 65 pN, it suffers a structural change and suddenly the molecule reaches an extension of about 1.7 times its natural length \cite{smith1996overstretching,cluzel1996overstretching}.  The nature of this overstretched DNA (which has been named S-DNA) stimulated a long debate in the last years \cite{dnamelting, new_melting}, as two different visions of the hydrogen bonds breaking could explain the transition \cite{newstatedna,cocco2004,Whitelam_SDNA}. Indeed, while it was clear that the double helix unwinds and the base pair align with the stretching force, it was not fully understood whether the B-S transition was caused by a complete melt into bubbles of the molecule (strand unpeeling) or rather a local bond breaking (a pure transition to a new state), until a coherent explanation was found \cite{Fu2010}. In this last work it is highlighted how both processes can happen according to experimental conditions, which can lead to either a fast transition to S-DNA or a single stranded molecule.\\
In order to understand better the mechanical properties of bio-molecules, it is essential to complement the experimental approach with theoretical models which are able to reproduce the main features of those systems. The simplest model that captures at least some basic mechanical features is the so-called Freely Jointed Chain (FJC) \cite{FJC}. This model depicts the molecule as a chain of non interacting units which  is stretched by an external force and where the monomers are discrete segments with an orientation in space. FJC model is qualitatively in agreement with the typical experimental measurements (e.g. in the regime of small forces), with some differences arising from the fact that pulled DNA extends even more than its contour length. In the intermediate regime of forces the FJC solution is however quantitatively different from the real elongation curves, showing then that the intermolecular interaction plays an important role.\\
A second model that  incorporate more features of real polymers  is the Worm Like Chain (WLC) \cite{kratky1949x}, the standard choice to fit the experimental measurements.  In this model, the molecule is seen as a whole continuous body, where the interaction is felt as a bending stiffness, which characterises the correlation at zero force. The WLC solution is a good improvement respect to the FJC, but an accurate comparison of the WLC result with real data still shows discrepancies between the model and actual systems \cite{elasticity}. One main difference comes from the fact that the fundamental building blocks of DNA behave extensibly, which results for high force into an elongated DNA longer than its native length at rest. This feature is then normally taken into account considering the building blocks  as springs and introducing a so called ``stretching modulus'' \cite{stretchingmod}. However this care is not enough, since the WLC can be a good approximation as long as the correlation length is much longer than the base pair length in DNA.\\
Some years ago Storm and Nelson \cite{storm-nelson} introduced the Discrete Persistent Chain (DPC) model, where both the coarse-grained nature (like in the FJC) and the bending stiffness (as in the WLC) of real molecules are incorporated. In this model the DNA pulling experiments are described using a Heisenberg-like Hamiltonian, where the stretching force is an external field and the interaction along the chain makes the polymer stiff. Their solution (where again a stretching modulus was considered) was in fair agreement with the experimental data of \cite{measures-Storm-nels} and lead to a fit which quantitatively improved the previous WLC results. Unfortunately, the author's nice  modelling is handicapped by an analytical treatment relying in a series of approximations which yielded unreasonable values as, for instance, the monomer's length being much smaller than the real base pair size for ss-DNA.\\
 It is important to note that all these are phenomenological models and to improve upon them one needs firstly to understand what they are actually modelling: what does  the discrete segment capture from the real DNA molecule? Does it model, for instance, a base pair or a bigger part of the DNA? In order to understand this model we revisit here the DPC model and fit it with experimental measurements.
\section{The Model}
\subsection{Single stranded DNA}
In the DPC  model \cite{storm-nelson,yanmarko,poland} the DNA is considered to be a chain  of $N$ interacting monomers with Hamiltonian
\begin{equation} 
\label{hamiltonian}
-\beta H(\hat{\bm{t}}) = f b_B \sum_{i=1}^N  \hat{t}_i \cdot \hat{z} + J_{B} \sum_{i=1}^{N-1}  \hat{t}_i \cdot \hat{t}_{i+1}
\end{equation}
where $\hat{\bm{t}}=\left(\tc_1, \ldots\tc_N\right)$, $f$  is the stretching force, $b_B$ the monomers' length and $J_B$ the interaction between the monomers, which tunes the bending stiffness of the chain. As we expect the segments to be aligned and to resist bending, $J_B$ is chosen as a ferromagnetic coupling. Vectors $\hz$ and $\tc_i$ lie on the unit sphere, the former pointing to the stretching direction, while the latter are aligned with the direction in space of the monomers. Note that the Hamiltonian  \eref{hamiltonian} is simply a Heisenberg model \cite{fisher1963} in external field $fb_B$. The free energy per monomer $F$ is
\begin{equation} \label{freeenergy}
-\beta F = \frac{1}{N}\log  \int_{S^2}\rmd^2 \hat{\bm{t}}\; e^{-\beta H(\bm{t})}\,,
\end{equation}
where the subscript $S^2$ denotes the integral over the unit sphere. As the fundamental blocs of real DNA are extensible themselves, the stretched DNA is longer than its contour length. To capture this feature we introduce a ``stretching modulus'' and rewrite the monomer's length as $b_B=b_B^{(0)}\left(1+f/Y\right)$ with $Y$ the Young modulus. The chain's elongation then reads
\begin{equation}
L\left(f;\bm{\mu}\right)=\frac{\partial (-\beta F) }{\partial f}= \frac{c_B}{N}\sum_{i=1}^N\bracket{\hat{t}_i \cdot \hat{z}}
\label{eq:elong}
\end{equation}
with $c_B=b_B^{(0)} \left(1+2f/Y\right)$, where $\bracket{\cdots}$ is the usual thermal average, and  $\bm{\mu}=(b^{(0)}_B,J_B,Y)$ are the parameters of the model. Note that the dependence of the elongation on the force is explicitly through $c_B$ and implicitly through the thermal average.\\
\subsection{Double stranded DNA}
The preceding model can be extended so as to capture the behaviour of double stranded B-DNA when pulled by strong forces. Since the first experiments by Smith {\em et al} \cite{smith1996overstretching} and Cluzel {\em et al} \cite{cluzel1996overstretching} it has been observed that double stranded B-DNA undergoes a very sharp transition when stretched by a force around 65 pN. When the pulling force reaches this critical value, the molecule extends its contour length of about 1.7 times in a very abrupt manner, showing a highly cooperational transition. The state of this over-stretched DNA is called S-DNA and its emergence has been tightly related to the experimental setup \cite{Fu2010}. To capture this we reconsider the Hamiltonian \eref{hamiltonian} and,  as in \cite{storm-nelson},  we allow the  monomers to be into two possible states: native B-DNA and denaturated S-DNA \cite{poland,storm-nelson,levine}. This yields the following Hamiltonian:
\begin{eqnarray}
\label{ising hamiltonian}
\hspace{-2cm}-\beta H \left(\bm{t}, \bsigma \right)& =& f\sum_{i=1}^N  b_{\sigma_i} \hat{t}_i\cdot\hat{z}+\gamma_B\sum_{i=1}^N \delta_{\sigma_i,B}+\sum_{i=1}^{N-1}  \left( J_{\sigma_i, \sigma_{i+1}} \hat{t}_i \cdot \hat{t}_{i+1}+ \epsilon_{\sigma_i, \sigma_{i+1}}\right)
\end{eqnarray}
where $\sigma_i\in\{B,S\}$ represents the state of monomer $i$. Here $b_{\sigma}$ represents the monomer's length in state $\sigma$ and $J_{\sigma,\sigma'}(=J_{\sigma',\sigma})$ the interaction between two neighbouring monomers in states $\sigma$ and $\sigma'$. In what follows we assume $\epsilon_{\sigma\sigma}=0$ and denote $J_{\sigma}=J_{\sigma\sigma}$. We also assume, as before, that the monomer's length of the native state is $b_B=b_B^{(0)}\left(1+f/Y\right)$, while the one in the denaturated state is rigid. The chain's elongation reads:
\begin{eqnarray}
L\left(f;\bm{\mu}\right)= \frac{c_B}{N}\sum_{i=1}^N\delta_{\sigma_{i},B}\bracket{\hat{t}_i \cdot \hat{z}}+\frac{b_S}{N}\sum_{i=1}^N\delta_{\sigma_{i},S}\bracket{\hat{t}_i \cdot \hat{z}}\,,
\end{eqnarray}
where the parameters of this second model are $\bm{\mu}=\{b^{(0)}_B,Y, b_{S}, J_B, J_{BS}, J_S, \gamma_{B}, \epsilon_{BS}\}$.

\section{ Parameter fitting}
The parameters  $\bm{\mu}$ can be fit to a set of experimental measurements $\{(f^{({\rm exp})}_{a},L^{({\rm exp})}_a)\}_{a=1}^{\mathcal{N}}$ by using Newton's method, viz.
\begin{equation}
\bm{\mu}^{(\ell+1)} =  \bm{\mu}^{(\ell)} - \bm{H} \cdot \nabla_{\bm{\mu}} \chi^2\,,\quad \ell=0,1,\ldots
\label{eq:newton}
\end{equation}
 Here $\bm{H}$ and $ \nabla_{\bm{\mu}} \chi^2$ represent the Hessian and gradient of the cost function
\begin{equation} \label{chi2}
\chi^2 = \sum_{a=1}^{\mathcal{N}} \left(L^{({\rm exp})}_a - L \left( f^{({\rm exp})}_a ;\bm{\mu}\right)\right)^2\,,
\end{equation}
respectively. They read as follows:
\begin{eqnarray}
\hspace{-1.5cm}\partial_{\mu_i} \chi^2 &=& -2\sum_{a=1}^{\mathcal{N}} \left(L^{({\rm exp})}_a - L \left( f^{({\rm exp})}_a ;\bm{\mu}\right)\right)\partial_{\mu_i}L \left( f^{({\rm exp})}_a ;\bm{\mu}\right)\,,\\
\hspace{-1.5cm}\partial^2_{\mu_j\mu_i} \chi^2 &=& 2\sum_{a=1}^{\mathcal{N}}\Bigg[\partial_{\mu_j}L \left( f^{({\rm exp})}_a ;\bm{\mu}\right) \partial_{\mu_i}L \left( f^{({\rm exp})}_a ;\bm{\mu}\right)\nonumber\\
&&\quad\quad -\left(L^{({\rm exp})}_a - L \left( f^{({\rm exp})}_a ;\bm{\mu}\right)\right)\partial^2_{\mu_j \mu_i}L \left( f^{({\rm exp})}_a ;\bm{\mu}\right)\Bigg]\,,\quad i,j=1,2,3\,.
\end{eqnarray}
As it is expected that $L^{({\rm exp})}_a \approx L \left( f^{({\rm exp})}_a ;\bm{\mu}\right)$, the second term in the Hessian is usually discarded and the latter is approximated by
\begin{eqnarray}
\partial^2_{\mu_j\mu_i} \chi^2 &\approx & 2\sum_{a=1}^{\mathcal{N}}\partial_{\mu_j}L \left( f^{({\rm exp})}_a ;\bm{\mu}\right) \partial_{\mu_i}L \left( f^{({\rm exp})}_a ;\bm{\mu}\right)\,.
\label{hessianapprox}
\end{eqnarray}
Note that the expressions for the gradient ${\bf \nabla}_{\bm{\mu}} \chi^2$ and the approximated Hessian $\bm{H}$ depend on  the derivatives of the elongation with respect to the parameters of the model, and these are simply correlation functions. For instance, in  model \eref{hamiltonian}, derivatives of $L$ with respect to $J_B$ and $b_B$ (derivatives with respect to $b^{(0)}_B$ and $Y$ follow from $b_B$ via the chain rule), yield the following correlation functions:
\begin{eqnarray}
\hspace{-1.5cm}\frac{\partial L }{\partial J_B}& = &  \frac{ c_B}{N}\sum_{i=1}^{N}\sum_{k=1}^{N-1}  \bracket{ \left(\tc_i \cdot \hz \right) \left(\tc_k \cdot \tc_{k+1} \right)}_{c}\,,\quad \frac{\partial L }{\partial b_{B}}=  \frac{f c_B}{N}\sum_{i,k=1}^{N} \bracket{ \left(\tc_i \cdot \hz\right) \left(\tc_k \cdot \hz\right)}_{c} \,,
\label{eq:correlationfunctions}
\end{eqnarray}
where $\bracket{\cdots}_{c}$ denotes connected correlation in the thermal average.\\
Thus in order to minimise efficiently the cost function $\chi^2$ by Newton's method, we need to find a way to deal with the thermal average appearing in eqs. (\ref{eq:elong},\ref{eq:correlationfunctions}) and the corresponding expressions for model \eref{ising hamiltonian}. Luckily the two models introduced are simple enough that, as we will see, a simple set of equations can be found to perform thermal averages efficiently.
\section{ Analytical solution with cavity method}
\subsection{Single stranded DNA}
Due to the one-dimensional nature of the problem, the solution of the DPC model is usually sought via the transfer matrix method, possibly expanding the eigenfunctions in spherical harmonics. However, while in the Heisenberg model with zero field the transfer matrix can be easily diagonalised with such an expansion \cite{takahashi1999heisenberg}, in this case the two terms in (\ref{hamiltonian}) do not commute and hence a variational method is generally adopted.  Another source of approximations is to consider the asymptotic behaviour of the elongation $L(f;\bm{\mu})$ for small and high forces and use the subsequent expressions (see \ref{appendix:smallforces}) to fit the data. As we want to fit the model in the whole range of forces to assess its validity, avoiding, in turn, unnecessary approximations, we proceed as follows: we use the cavity method \cite{mezardparisi} to write exact solutions for the constrained local partition function and the chain's elongation; then we evaluate the correlation functions \eref{eq:correlationfunctions} involved in the minimisation of the  cost function  \eref{chi2}  by writing down equations for the propagation of a perturbation in the cavity equations.\\
Let us assume that the chain is very long, so as to neglect boundary effects and consider a monomer with variable $\hat{t}_0$ in the bulk. The Hamiltonian is then rewritten as
\begin{eqnarray}
-\beta H(\hat{\bm{t}})&=&f b_B\hat{t}_0 \cdot \hat{z}+J_{B}\hat{t}_0\cdot\sum_{i\in\partial0}  \hat{t}_{i}-\beta H^{(0)}(\hat{\bm{t}})
\end{eqnarray}
where $-\beta H^{(0)}(\hat{\bm{t}})$ stands for the system without the monomer $\hat{t}_0$ and $\partial0$ denotes the set of sites neighbouring the site $0$. The constrained partition function $Z \left(\tc_0 \right)$ at that site in the bulk reads
\begin{eqnarray}
Z \left(\tc_0 \right)&=&e^{f b_B\hat{t}_0 \cdot \hat{z}}\int_{S^2}\rmd^2 \hat{\bm{t}}_{\partial 0}e^{J_{B}\hat{t}_0\cdot\sum_{i\in\partial0}  \hat{t}_{i}}Q(\hat{\bm{t}}_{\partial 0})
\end{eqnarray}
with $ \hat{\bm{t}}_{\partial 0}=\{\tc_i|i\in\partial 0\}$ and  where we have defined
\begin{equation}
\hspace{-1cm}Q(\hat{\bm{t}}_{\partial 0})=\int_{S^2}\rmd^2 \hat{\bm{t}}_{\backslash (0\cup \partial 0)}\; e^{-\beta H^{(0)}(\hat{\bm{t}})}\,,\quad\quad \quad\hat{\bm{t}}_{\backslash (0\cup \partial 0)}=\{\tc_{i}|  i\not\in 0\cup \partial 0\}\\
\end{equation}
as the cavity (i.e. in the absence of site $0$) partititon function. The latter obviously factorises for open chains, or for very long closed ones, having $Q(\hat{\bm{t}}_{\partial 0})=\prod_{i\in\partial 0}Q(\tc_i)$. This allows us to write the following expressions for the  partition function $Z \left(\tc \right)$ in terms of its cavity counterpart $Q \left(\tc\right)$ in the bulk, viz:
\begin{eqnarray}
Q \left(\tc\right) &=& e^{fb_B \tc \cdot \hz}\int_{S^2}\rmd^2 \hu \; e^{J_B \tc \cdot \hu}Q \left(\hu\right)\label{eq:cav1}\\
Z \left(\tc \right) &=& e^{fb_B \tc \cdot \hz} \left(\int_{S^2} \rmd^2 \hu \; e^{J_B \tc \cdot \hu}Q\left(\hu\right)\right)^2\label{eq:mar1}\,.
\end{eqnarray}
The expression  \eref{eq:elong} for the elongation $L(f)$ simply becomes
\begin{equation}
\label{elasticelongation}
L \left(f\right)= c_B \int_{S^2}  \rmd^2 \tc  \; P\left(\tc\right) \left(\tc \cdot \hz\right)\,,
\end{equation}
in which $P\left(\tc\right)=Z\left(\tc\right)/Z$ and $Z=  \int_{S^2} \rmd^2 \tc \; Z \left(\tc \right)$ is a normalisation constant. A numerical solution of $Q\left(\tc\right)$ can be found by simple iteration of eq. \eref{eq:cav1}. Once $Q\left(\tc\right)$ is known we can calculate  $Z \left(\tc \right)$ and $L(f)$ by using equations \eref{eq:mar1} and \eref{elasticelongation}, respectively.\\
In order to fit the experimental measurements we need to calculate the correlation functions appearing in \eref{eq:correlationfunctions} efficiently. In the cavity method this boils down to calculating  the perturbations of $P\left(\tc\right)$ with respect to $J_B$ and $b_B$. Indeed, by comparing \eref{eq:correlationfunctions} and \eref{elasticelongation} we note that
\begin{eqnarray}
\frac{1}{N}\sum_{i=1}^{N}\sum_{k=1}^{N-1}  \bracket{ \left(\tc_i \cdot \hz \right) \left(\tc_k \cdot \tc_{k+1} \right)}_{c}&=&\int_{S^2}  \rmd^2 \tc  \; \frac{\partial  P\left(\tc\right)}{\partial J_B}  \left(\tc \cdot \hz\right)\label{eq:corr1}\\
\frac{f }{N}\sum_{i,k=1}^{N} \bracket{ \left(\tc_i \cdot \hz\right) \left(\tc_k \cdot \hz\right)}_{c} &=& \int_{S^2}  \rmd^2 \tc  \;\frac{\partial  P\left(\tc\right) }{\partial b_{B}} \left(\tc \cdot \hz\right)\label{eq:corr2}
\end{eqnarray}
where the derivatives of $P(\tc)$ can be expressed as follows:
\begin{equation}
\frac{\partial P\left(\tc\right)}{\partial x}=\frac{1}{Z} \frac{\partial Z\left(\tc\right)}{\partial x} -P\left(\tc\right)\frac{1}{Z} \frac{\partial Z}{\partial x}\,\quad\quad\quad x=J_B,b_B\,.
\label{eq:pertP}
\end{equation} 
As $Z \left(\tc \right)$ and $Q\left(\hu\right)$ are related via equation \eref{eq:mar1}, perturbations of $Z(\hat{t})$ correspond to perturbations of $Q\left(\hu\right)$, \textit{viz.}
\begin{eqnarray}
\frac{\partial Z \left(\tc \right)}{\partial J_B} &=&2e^{-fb_B \tc \cdot \hz} Q \left(\tc\right)\frac{\partial Q \left(\tc\right)}{\partial J_{B}}\label{eq:pertZ1} \\
\frac{\partial Z \left(\tc \right)}{\partial b_B} &=&-f\left( \tc \cdot \hz\right)Z \left(\tc\right)+2 e^{-fb_B \tc \cdot \hz} Q \left(\tc\right)\frac{\partial Q \left(\tc\right)}{\partial b_B}\label{eq:pertZ2}
\end{eqnarray}
with similar equations for the normalisation constant $Z$:
\begin{eqnarray}
\frac{\partial Z}{\partial J_{B}}&=&   \int_{S^2} \rmd^2 \tc 2e^{-fb_B \tc \cdot \hz} Q \left(\tc\right)\frac{\partial Q \left(\tc\right)}{\partial J_{B}}\label{eq:pertnormZ1}\\
\frac{\partial Z}{\partial b_{B}}&=& \int_{S^2} \rmd^2 \tc \left[-f\left( \tc \cdot \hz\right)Z \left(\tc\right)+2 e^{-fb_B \tc \cdot \hz} Q \left(\tc\right)\frac{\partial Q \left(\tc\right)}{\partial b_B}\right]\label{eq:pertnormZ2}\,.
\end{eqnarray}
Perturbations of $Q\left(\hu\right)$ obey the following equations
\begin{eqnarray}
\frac{\partial Q \left(\tc\right)}{\partial J_{B}} &=& e^{fb_B \tc \cdot \hz}\int_{S^2}\rmd^2 \hu\;e^{J_B \tc \cdot \hu}\left[ \;\left(\tc \cdot \hu\right) Q \left(\hu\right)+\frac{\partial Q \left(\hu\right)}{\partial J_{B}} \right]\label{eq:pertQ1}\\
\frac{\partial Q \left(\tc\right)}{\partial b_{B}} &=& f\left(\tc \cdot \hz \right)Q \left(\tc\right)+ e^{fb_B \tc \cdot \hz}\int_{S^2}\rmd^2 \hu \; e^{J_B \tc \cdot \hu}\frac{\partial Q \left(\hu\right)}{\partial b_{B}}\label{eq:pertQ2}
\end{eqnarray}
To calculate the correlations one proceeds as follows: the set equations (\ref{eq:cav1},\ref{eq:pertQ1},\ref{eq:pertQ2}) for $\{Q(\tc),\partial_{J_{B}} Q \left(\tc\right),\partial_{b_B} Q \left(\tc\right)\}$ is solved numerically. Their solution is then used to calculate the perturbations of $Z \left(\tc \right)$ and its normalising constant $Z$, given by eqs (\ref{eq:pertZ1}-\ref{eq:pertnormZ2}). These are subsequently used in \eref{eq:pertP} to obtain the perturbations of $P(\tc)$ which are finally plugged in eqs. (\ref{eq:corr1},\ref{eq:corr2}) to obtain the correlations. \\
Note that is possible  to show that one can recover the lhs of eqs. (\ref{eq:corr1},\ref{eq:corr2}) from the rhs. To do this one only needs to write down an explicit solution of $\partial_x Q \left(\tc\right)$ by formally solving its corresponding equation via infinite iteration. Indeed, to illustrate let us consider the perturbation with respect to $b_B$. The solution of equation \eref{eq:pertQ2} can be written as follows
\begin{eqnarray}
\hspace{-2cm}\frac{\partial Q \left(\tc_i\right)}{\partial b_{B}}&=&\left(\tc_i \cdot \hz \right)Q \left(\tc_i\right)\\
\hspace{-2cm}&+&f\sum_{k=0}^\infty\int_{S^2}\left[\prod_{\ell=0}^k\rmd^2 \tc_{i+1+\ell}\right] e^{\sum_{\ell=0}^k\left(fb_B \tc_{i+\ell} \cdot \hz+J_B \tc_{i+\ell} \cdot \tc_{i+\ell+1}\right)}  \left(\tc_{i+k+1} \cdot \hz \right)Q \left(\tc_{i+k+1}\right)\nonumber
\end{eqnarray}
This, in turn, yields:
\begin{eqnarray}
\hspace{-2cm}\frac{\partial Z \left(\tc_i \right)}{\partial b_B} &=&f\left( \tc_i \cdot \hz\right)Z \left(\tc_i\right)+2f \sum_{k=0}^\infty\int_{S^2}\left[\prod_{\ell=0}^k\rmd^2 \tc_{i+1+\ell}\right]Z(\tc_i,\tc_{i+1},\ldots \tc_{i+k+1})\left(\tc_{i+k+1} \cdot \hz \right)\nonumber
\end{eqnarray}
where  $Z(\tc_i,\tc_{i+1},\ldots \tc_{i+k+1})$ is the constrained partition function of a segment of the chain of length $k+1$:
\begin{eqnarray}
\hspace{-1cm} Z(\tc_i,\tc_{i+1},\ldots \tc_{i+k+1})= Q\left(\tc_i\right)e^{\sum_{\ell=1}^k fb_B \tc_{i+\ell} \cdot \hz+J_B\sum_{\ell=0}^k \tc_{i+\ell} \cdot \tc_{i+\ell+1}}  Q \left(\tc_{i+k+1}\right)
\end{eqnarray}
Note that using equation \eref{eq:cav1} we have that:
\begin{equation}
\int_{S^2}\rmd^2 \tc_{i} Z(\tc_i,\tc_{i+1},\ldots \tc_{i+k+1})=Z(\tc_{i+1},\tc_{i+2},\ldots \tc_{i+k+1})\,.
\end{equation}
This implies that the normalisation of $Z(\tc_i,\tc_{i+1},\ldots \tc_{i+k+1})$ is precisely $Z$ and we can write $P(\tc_i,\tc_{i+1},\ldots \tc_{i+k+1})=Z(\tc_i,\tc_{i+1},\ldots \tc_{i+k+1})/Z$.  These results yield the following nice and compact expression  for $\partial_{b_{B}}P(\tc)$:
\begin{eqnarray}
\hspace{-2.5cm}\frac{\partial P\left(\tc_i\right)}{\partial b_{B}}&&=fP \left(\tc_i\right)\left[ \tc_i \cdot \hz-\bracket{ \tc_i \cdot \hz}\right]\\
\hspace{-2.5cm}&&+ 2f \sum_{k=0}^\infty\int_{S^2}\left(\prod_{\ell=0}^k\rmd^2 \tc_{i+1+\ell}\right)P(\tc_i,\tc_{i+1},\ldots \tc_{i+k+1})\left\{\tc_{i+k+1} \cdot \hz  - \bracket{\tc_{i+k+1} \cdot \hz}\right\} \nonumber
\end{eqnarray} 
which can be used to nicely recover the correlation function appearing on the lhs of eq. \ref{eq:corr2}. These preceding expressions are however impractical and it is preferable to solve the aforementioned equations numerically.
\subsection{Double stranded DNA}
This model is solved as before using the cavity method. Although the expressions in this case are a bit more involved, we do not linger in minor details and simply report the most revelant equations for the curious reader. As in this case we have two states per monomer we obtain equations for the constrained partition function $ Z_{\sigma}\left(\tc\right)$ and its cavity counterparts $ Q_{\sigma}\left(\tc\right)$:
\begin{eqnarray}
 \label{ising real density}
 Q_{\sigma}\left(\tc\right)&=&e^{\gamma_B\delta_{\sigma,B}+f b_\sigma  \tc\cdot \hat{z}} \sum_{\tau\in\{B,S\}} e^{\varepsilon_{\sigma\tau}} \int_{S^2} \rmd^2 \hu \; e^{J_{\sigma \tau} \tc\cdot\hu } Q_\tau\left(\hu \right)\\
 Z_{\sigma}\left(\tc\right)&=&e^{\gamma_B\delta_{\sigma,B}+f b_\sigma  \tc\cdot \hat{z}} \left(\sum_{\tau\in\{B,S\}} e^{\varepsilon_{\sigma\tau}} \int_{S^2} \rmd^2 \hu \; e^{J_{\sigma \tau} \tc\cdot\hu } Q_\tau\left(\hu \right)\right)^2
\end{eqnarray}
with $\sigma\in\{B,S\}$. The elongation $L$ takes the simple form:
\begin{equation}
 \label{ising elong}
L\left(f\right)=\;\int_{S^2} \rmd^2 \tc\,\left(\tc \cdot \hz \right)\left[  c_BP_B(\tc)+b_{S}  P_S(\tc)\right]
\end{equation}
where $P_\sigma\left(\tc\right)=Z_\sigma\left(\tc\right)/Z$ with $Z= \sum_\sigma \int_{S^2} \rmd^2 \tc \; Z_\sigma \left(\tc \right)$ a normalisation constant.  The expressions for the propagation of pertubations in this case are fairly tedious and we prefer to report their exact expression somewhere else \cite{FrancescoThesis}.

\section{The fitting with experimental measurements and Monte Carlo Simulations}
In this section we summarise the results of our fit to experimental measurements provided by the ``Small Biosystems Laboratory'' in Barcelona, both for single stranded and double stranded DNA. To check the validity of our analytical results we have also performed Monte Carlo simulations of both Hamiltonians (\ref{hamiltonian},\ref{ising hamiltonian}) using Heat Bath \cite{heatbath} for the vector variables (combined with Metropolis for the discrete variables in the second model) for a system with $N=1000$ monomers and open boundary condition.
Note that we do  Monte Carlo simulations using the values of already fitted parameters so as to check the correctness of our work. Using Monte Carlo simulation for fitting the experimental data is a very time-consuming and impractical task.
\subsection{Single stranded DNA}
\begin{figure}[t]
\includegraphics[width=0.8\columnwidth]{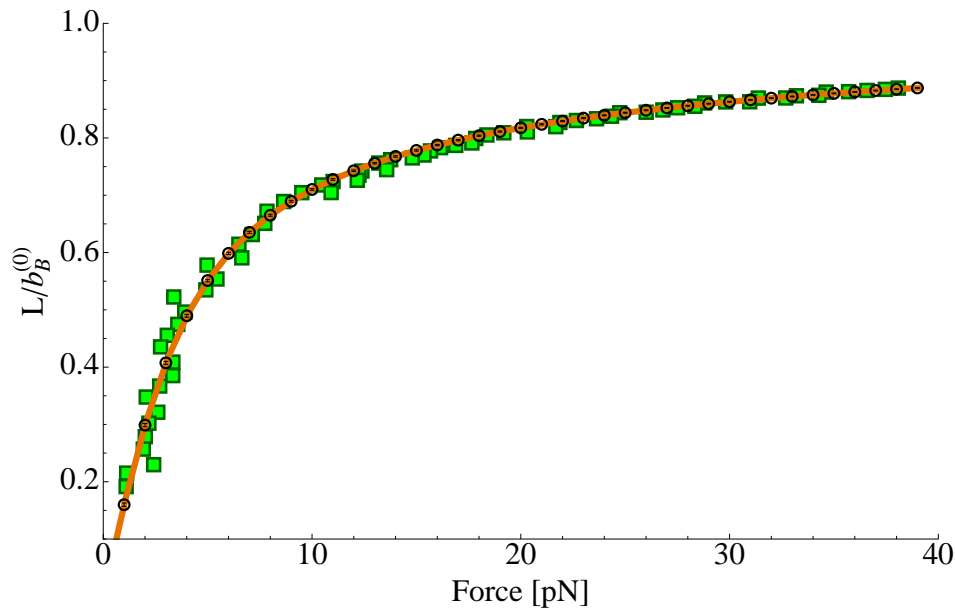}
\caption{ Fit of ss-DNA stretching measurements taken at the ``Small Biosystems Lab'' in Barcelona, 100 experimental points..  Experimental data is reported in full squares, theoretical solution in straight line while open circles are Monte Carlo simulations. Values of the fit: $b^{(0)}_{B}=0.63$ nm, $J_{B}=7.68$ pN nm, $\chi^2=1.3\times 10^{-4}$. We take room temperature $k_B T=4.04738$ pN nm.}
\label{fit 00}
\end{figure}
We first proceed to fit model \eref{hamiltonian} to a set of experimental measurements of ss-DNA stretching. The results are reported in Figure \ref{fit 00}, where we plot experimental results, the fit and Monte Carlo simulations. The experiments were performed  in a 10 mM NaCl solution on a chain of 3000 bases whose distance is estimated to approximately 0.7 nm by X ray diffraction. Because of the reduced range of forces (from about 0 to 40 pN) the single monomers' extension is not detectable and therefore we keep the inverse Young modulus fixed to zero during the fitting. Leaving $Y^{-1}$ free would indeed return a small value, thus this choice allows us to reduce the number of fitting parameters. The results of the fit are a monomer's length  $b_B^{(0)} \simeq$ 0.63 nm and a coupling constant $J_R \simeq 7.68$ pN nm, while for cost function we obtain $\chi^2=1.3\times10^{-4}$. This result suggests the model's monomer corresponds to the actual DNA base distance, at least for the single stranded chain. This is very encouraging, since no discrete model so far could capture the real scaling of the molecule. As stated before, in fact, the absence of a bending resistance results in a monomer's length longer than the actual one (the Kuhn segment), while the DPC model yielded a ss-DNA monomer's length much smaller than the base distance \cite{storm-nelson}. This means that the qualitative picture of the model can give  meaningful and expendible quantitative results if a rigorous analysis of the problem is carried out. This last point is also confirmed by the more than good agreement between our analytic solution and the Monte Carlo simulation.
\subsection{Double stranded DNA}
\begin{figure}[t]
\includegraphics[width=0.5\columnwidth]{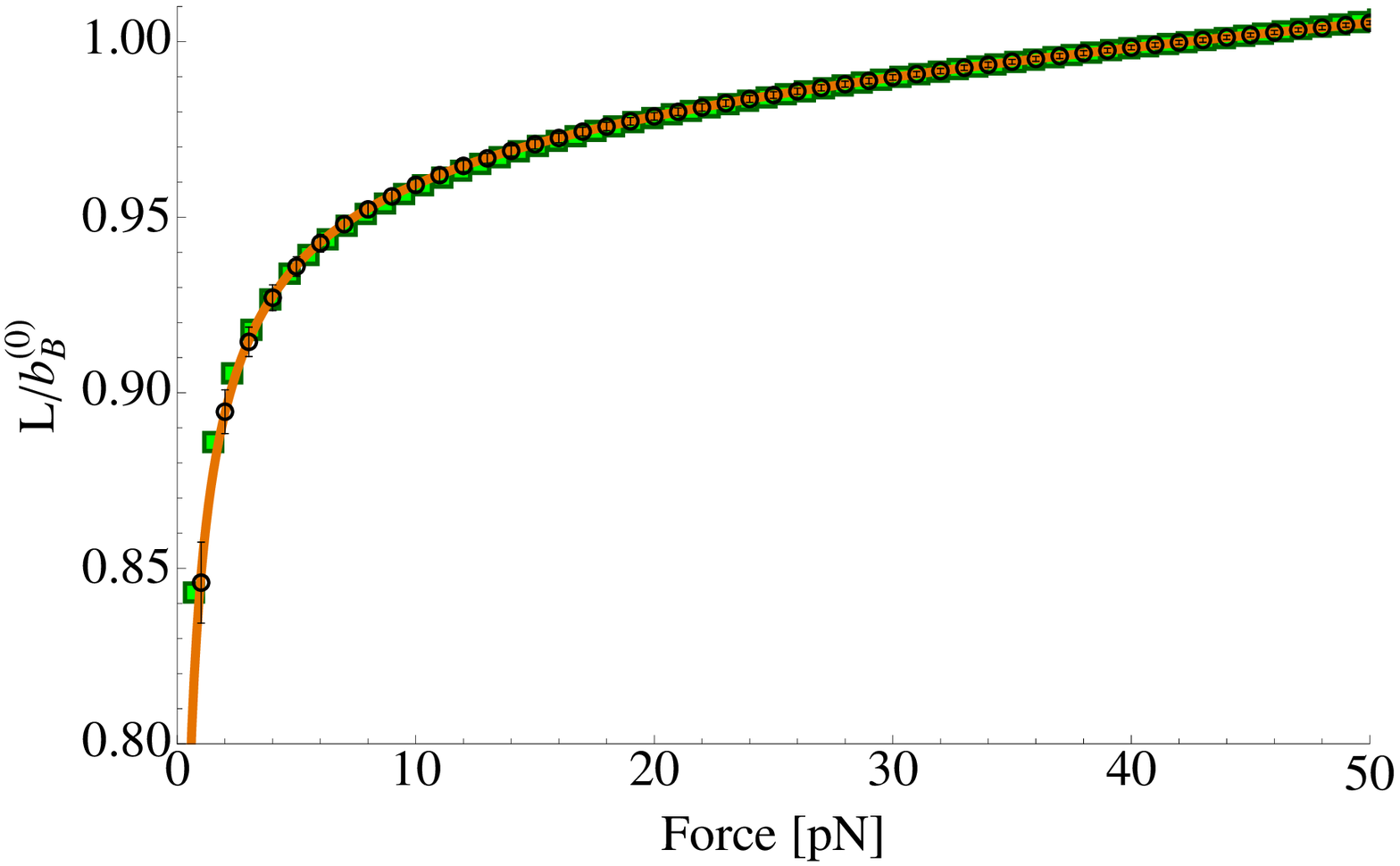}\includegraphics[width=0.5\columnwidth]{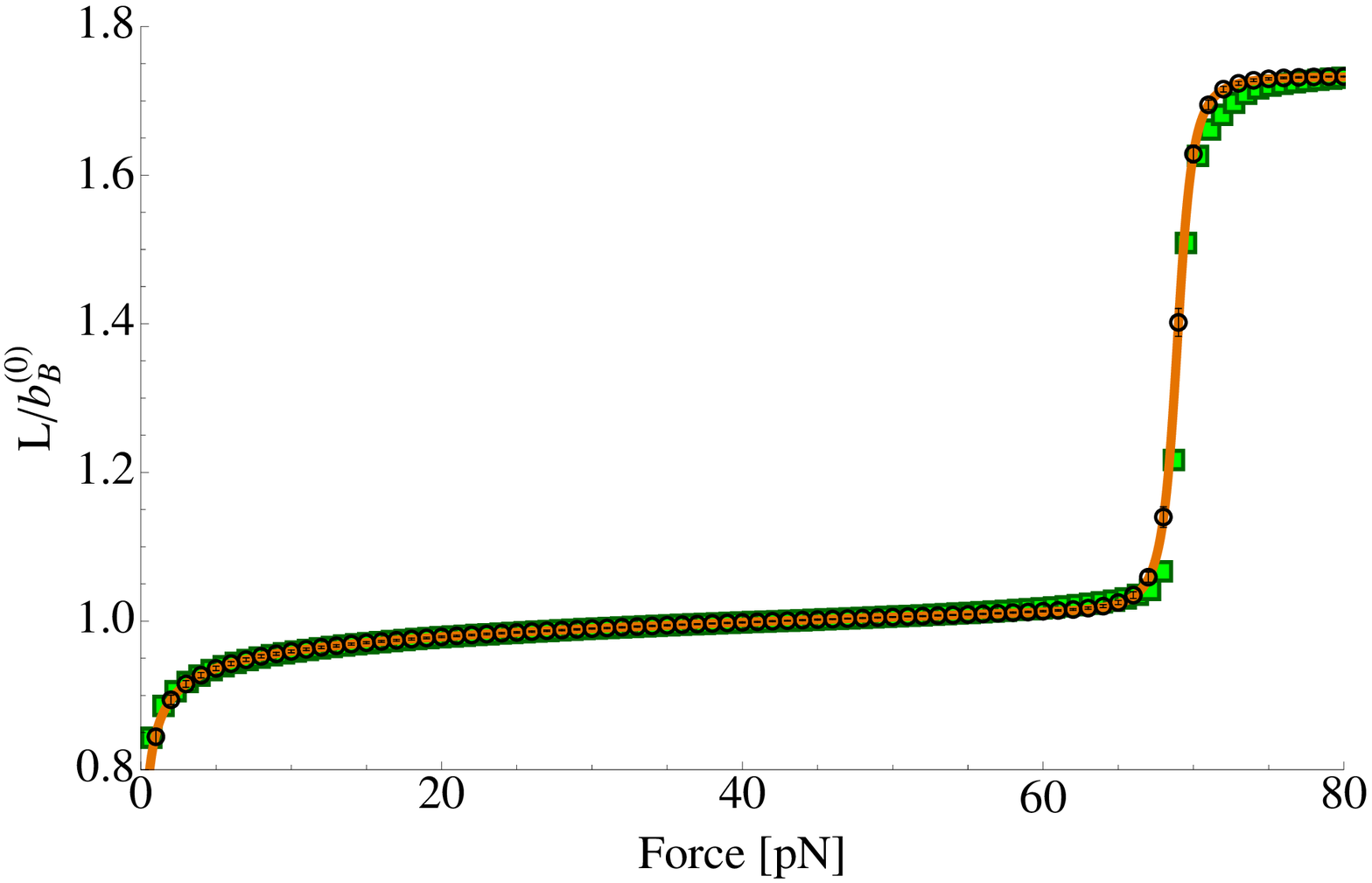}
\caption{Left figure:  Fit of ds-DNA stretching measurements taken at the ``Small Biosystems Lab'' in Barcelona, 100 experimental points. Experimental data is reported in full squares, theoretical solution in straight line while open circles are Monte Carlo simulations. The force range is narrow to avoid any effect linked to the overstretching from above 50 pN and to thermal fluctuation from below 4 pN. Values of the fit: $b^{(0)}_B=3.22$ nm, $Y=4570.38$ pN, $J_B=56.26$ pN nm, $\chi^2=9\times 10^{-5}$. Here we have assumed room temperature $k_B T=4.04738$ pN nm. Right figure: Fit of overstretched DNA. Experimental points are plotted in filled squares, theoretical solution in straight line and Monte Carlo simulations in open circles. The data shows the typical behavior of dsDNA stretched by high forces: the double plateau and the jump around 65 pN are easily recognizable. The fit is operated over 100 points, starting from $f=1.5$ pN. The parameters values returned by the fit are reported in table \ref{parameters fit}, $\chi^2 = 4\times 10^{-4}$.}
\label{fit 0}
\end{figure}

Figure \ref{fit 0} summarises the results of our fit to experimental measurements for ds-DNA. For sake of efficiency, the fitting is done in two stages. Firstly, we fit the first model to to experimental measurements with forces below 50 pN, that is, below the overstretching transition (left figure in \ref{fit 0}).  The agreement is beyond any doubt, finding  the monomers' length of around $b^{(0)}_B=3.22$ nm, the Young modulus to be $Y=4570.38$ pN and $J_B=56.26$ pN nm.
We then proceed to fit the second model to the set of experimental measurements up to range of forces to 80 pN, while keeping the values of $b_B^{(0)}$, $J_B$ and $Y$ fixed to the previous fitted values. The results of the fit, together with results from Monte Carlo simulations are shown in Figure \ref{fit 0} (right figure) and the parameters are reported in Table \ref{parameters fit}. The result is excellent, proving that this approach can realistically approximate the physical system: as already mentioned, we find a native monomers' length which is of the order of 10 base pairs (the latter being $\sim$ 3.3 \AA), which interestingly corresponds to the DNA helix period. This is much smaller than what is achieved with the FJC and somehow better than what is obtained in \cite{storm-nelson}, since the very small value (less than a base pair) calculated therein has a difficult physical interpretation. The denaturated unit length $b_B$  is 1.75 times bigger than the native one, giving back the known ratio between S-DNA and B-DNA total length: this value confirms that the helix unwinds, even if it doesn't give any insight on the melting into bubbles. Finally we get from the values of $J_B$, $J_{BS}$ and $J_S$ that the native strand is much stiffer (and so more correlated as well) than the denaturated portion of the molecule.
\begin{table}
\begin{center}
\begin{tabular}{|c|c|c|c|c|c|c|}
\hline
$b^{(0)}_B$ & $b_{S}$ & $J_B$ & $J_{BS}$ & $J_S$ & $\gamma_{B}$ & $\epsilon_{BS}$ \\
\hline
3.22 & 5.63 &56.26  & 26.35& 0.81& 106.61 & -2.75  \\
\hline
\end{tabular}
\caption{Parameters of the fit. For the sake of compactness we write $J_{\sigma \sigma} \equiv J_{\sigma}$. The parameters $b_{\sigma}$ are given in nm while $J_{\sigma\sigma'}$ and $\epsilon_{\sigma\sigma'}$  are given in pN nm.}
\label{parameters fit}
\end{center}
\end{table}

\section{Conclusions}
In this paper we have revisited the very simple phenomenological model of Storm and Nelson which faithfully reproduces the mechanical properties of dsDNA and where the monomer's length represents one helix period of the DNA polymer. The model is treated with the cavity method which facilities the fitting by easily allowing to understand changes of the fitting parameters as perturbations of the cavity equations.\\
We believe that the good analytic and numerical results obtained in this paper can be the starting point of many further applications in biophysics and molecular biology, particularly in heterogeneous polymers, block copolymers and protein helix-coil transitions in general. We also think that more realistic yet solvable models can be drawn from this study. For instance, since we uncover that a DPC's monomer corresponds to a whole helix period, we could incorporate its internal degrees of freedom in such a way to study the single base pair's behaviour.

\section*{Acknowledgements}
The authors would like to thank Felix Ritort, Joan Camunas and Josep Maria Huguet for providing the experimental measurements carried out at the ``Small Biosystems lab'' in Barcelona and for discussions.

\section*{References}
\bibliographystyle{iopart-num}
\bibliography{bibliography}

\providecommand{\newblock}{}
\begin{thebibliography}{10}
\expandafter\ifx\csname url\endcsname\relax
  \def\url#1{{\tt #1}}\fi
\expandafter\ifx\csname urlprefix\endcsname\relax\def\urlprefix{URL }\fi
\providecommand{\eprint}[2][]{\url{#2}}

\bibitem{bustamante1992}
Smith S, Finzi L and Bustamante C 1992 {\em Science\/} {\bf 258} 1122

\bibitem{elasticity}
Bustamante C, Marko J, Siggia E and Smith S 1994 {\em Science\/} {\bf 265} 1599

\bibitem{bendingstiff}
Hagerman P 1988 {\em Annu. Rev. Biophys. Biophys. Chem.\/} {\bf 17} 265--286

\bibitem{smith1996overstretching}
Smith S, Cui Y and Bustamante C 1996 {\em Science\/} {\bf 271} 795--799

\bibitem{cluzel1996overstretching}
Cluzel P {\em et~al.\/} 1996 {\em Science\/} {\bf 271} 792

\bibitem{dnamelting}
Williams M, Rouzina I and Bloomfield V 2002 {\em Acc. Chem. Res.\/} {\bf 35}
  159--166

\bibitem{new_melting}
van Mameren J {\em et~al.\/} 2009 {\em Proc. Natl. Acad. Sci. USA\/} {\bf 106}
  18231

\bibitem{newstatedna}
Lebrun A and Lavery R 1996 {\em Nucleic Acids Res.\/} {\bf 24} 2260

\bibitem{cocco2004}
Cocco S {\em et~al.\/} 2004 {\em Phys. Rev. E\/} {\bf 70} 011910

\bibitem{Whitelam_SDNA}
{Whitelam} S, {Geissler} P~L and {Pronk} S 2010 {\em ArXiv e-prints\/}
  (\textit{Preprint} \eprint{1002.2169})

\bibitem{Fu2010}
Fu H, Chen H, Marko J and Yan J 2010 {\em Nucleic Acids Research\/} ISSN
  0305-1048

\bibitem{FJC}
Flory P 1969 {\em {Statistical mechanics of chain molecules}\/} (Interscience,
  New York)

\bibitem{kratky1949x}
Kratky O and Porod G 1949 {\em Recl. Trav. Chim. Pays Bas\/} {\bf 68}
  1106--1122

\bibitem{stretchingmod}
Hegner M, Smith S and Bustamante C 1999 {\em Proc. Natl. Acad. Sci. USA\/} {\bf
  96} 10109

\bibitem{storm-nelson}
Storm C and Nelson P~C 2003 {\em Phys. Rev. E\/} {\bf 67} 051906

\bibitem{measures-Storm-nels}
Rief M, Clausen-Schaumann H and Gaub H 1999 {\em Nat. Struct. Biol.\/} {\bf 6}
  346--350

\bibitem{yanmarko}
Yan J and Marko J 2003 {\em Phys. Rev. E\/} {\bf 68} 11905

\bibitem{poland}
Poland D and Scheraga H~A 1966 {\em J. Chem. Phys.\/} {\bf 45} 1456--1463

\bibitem{fisher1963}
Fisher M~E 1964 {\em Am. J. Phys.\/} {\bf 32} 343--346

\bibitem{levine}
Chakrabarti B and Levine A~J 2005 {\em Phys. Rev. E\/} {\bf 71} 031905

\bibitem{takahashi1999heisenberg}
Takahashi M 1999 {\em {Thermodynamics of one-dimensional solvable models}\/}
  (Cambridge Univ Pr)

\bibitem{mezardparisi}
Mezard M and Parisi G 2001 {\em Eur. Phys. J. B\/} {\bf 20} 217--233

\bibitem{FrancescoThesis}
Massucci F~A in preparation {\em in preparation\/} Ph.D. thesis King's College
  London

\bibitem{heatbath}
Miyatake Y {\em et~al.\/} 1986 {\em J. Phys. C Solid State\/} {\bf 19}
  2539--2546

\end{thebibliography}

\appendix

\section{Expansion for small forces}
\label{appendix:smallforces}
Looking at the expression \eref{eq:elong} for $L(f)$, we note that the dependence on the force appears explicitly through the definition of $c_B$ and implicitly through the thermal average of  $\bracket{\hat{t}_i\cdot\hat{z}}$. Expanding the latter in powers of $f$ we write:
\begin{eqnarray}
\hspace{-2cm}\bracket{\hat{t}_i\cdot\hat{z}}&=&\bracket{\hat{t}_i\cdot\hat{z}}_{f=0}+f\left(\frac{d}{df}\bracket{\hat{t}_i\cdot\hat{z}}\right)_{f=0}+\frac{1}{2}f^2\left(\frac{d^2}{df^2}\bracket{\hat{t}_i\cdot\hat{z}}\right)_{f=0}+\mathcal{O}\left(f^3\right)\,.
\end{eqnarray}
Noting that $\bracket{\hat{t}_i\cdot\hat{z}}_{f=0}$ is identically zero (as this is Heisenberg model in one dimension at zero external field) and keeping the lowest terms in $f$ we can write:
\begin{equation}
L=b_B^{(0)}f \chi+\mathcal{O}(f^2)
\label{app:Lf}
\end{equation}
with 
\begin{equation}
\chi=\frac{1}{N}\sum_{i,k=1}^N\bracket{(\hat{t}_i\cdot\hat{z})(\hat{t}_k\cdot\hat{z})}_{f=0}
\end{equation}
being the magnetic susceptibility at zero external field. Assuming periodic boundary conditions $\chi$ takes the following form (see, for instance, \cite{takahashi1999heisenberg}):
\begin{equation}
\chi=\frac{1}{3}\frac{J_{B}+J_B~\cotanh(J_B)-1}{J_{B}-J_B~\cotanh(J_B)+1}\,.
\label{app:sus}
\end{equation}
Note that eqs. \eref{app:Lf} and \eref{app:sus} give an explicit expression of $L(f,b_B^{(0)},J_B)$, amenable for fitting $b_B^{(0)}$ and $J_B$ to experimental measurements but only valid in the linear regime of forces.\\
A similar analysis can be also done in the region for large forces by expanding in powers of $J_B$ instead. This is a pedagogical exercise left to the reader.

\end{document}